# Strain and voltage control of magnetic and electric properties of FeRh films

**Ignasi Fina and Josep Fontcuberta**

Institut de Ciència de Materials de Barcelona (ICMAB-CSIC), Campus UAB, Bellaterra, 08193, Catalonia, Spain

E-mail: ifina@icmab.es, fontcuberta@icmab.cat

**Abstract**

FeRh based alloys may display an uncommon transition from a ferromagnetic to an antiferromagnetic state upon cooling. The transition takes place roughly above room temperature and it can be sensitively modulated by composition and external parameters, including pressure and strain. Consequently, thin films of FeRh have received attention for applications in spintronics, antiferromagnetic spintronics and sensing. Interestingly, the extreme sensitivity of its properties to strain has created expectations for energy friendly voltage-control of the magnetic state of FeRh, with a number of potential applications at the horizon. Here, after summarizing the current understanding of strain effects on the magnetic properties of FeRh thin films, we review achievements on exploiting piezoelectric substrates for in-operando tuning of their magneto-electric properties. We end with a brief summary and an outlook for future initiatives.

## 1. Introduction

Materials at the verge of phase transitions offer the highest sensitivity to appropriate external stimuli and have found applications in advance technologies, including data storage, thermal storage or sensing, to mention few. In some magnetic materials, spins can be intimately linked to orbital and electronic degrees of freedom and in those materials, magnetic transitions are almost synonyms of magnetostructural phase transitions. The energy balance between competing phases can be tailored by temperature, magnetic field, pressure, etc. and thus an extremely rich variety of responses can be obtained. Alloys within the binary Fe-Rh system, are known since the late 1930s [1] to be one of such materials. Close to the equiatomic composition, FeRh displays an uncommon ferromagnetic (FM) to antiferromagnetic (AFM) transition at a temperature $T^* \approx 350$ K. At higher temperature, FeRh is paramagnetic with Curie temperature $T_C \approx 670$ K. Interestingly the AFM-FM transition occurs at a temperature $T^*$ that is near room temperature and it can be conveniently tailored by adjusting the Fe:Rh composition or by alloying with other metals. The AFM-FM transition is $1^{st}$ order, and it is accompanied by a significant volume expansion of about 1%, [2] thus justifying the label of "magnetostructural" material. In bulk form, the FM-to-AFM and AFM to FM transitions are sharp and hysteretic, as typical of $1^{st}$ order transitions. From now on, we define $T^*$ as the average transition temperature while increasing or decreasing temperature. Interestingly, the crystal symmetry is preserved across the transition (cubic, CsCl type) [3]. Not surprisingly, under hydrostatic pressure (P), $T^*$ is shifted to higher temperatures ($dT^*/dP = 4.33$ K/kbar) [4] reflecting the fact that the unit cell of the AFM phase is smaller than that of the FM one. Oppositely, an applied magnetic field (H), shifts $T^*$ to lower values ($dT^*/dH = -0.8$ K/kOe [5]). The magnetic ordering of the FM and AFM phases is known since 1960s [3]. The AFM phase is a G-type, which implies a FM ordering of spins within the (111) plane and AFM ordering between (111) planes. Whereas in the AFM phase only the Fe ions carry magnetic moment (about 3 $\mu_B$), in the FM phase both Fe and Rh ions carry moment (3.2 $\mu_B$ and 0.9 $\mu_B$) respectively [6]. The electric resistivity of the FM phase is smaller than that of the AFM and thus the resistivity displays an abrupt drop (by about 45%) upon crossing $T^*$ [7,8]. It was also early recognized that the transition at $T^*$ is accompanied by a large entropy change [8,9]. Whereas the phenomenology of the magnetic phase transition in FeRh, including its sensitivity to fine composition modulation, is well established, the mechanisms that drive it, remain unsolved and the relative weight of the structural changes and electronic reordering have been much discussed. See Lewis et al. [5] for an extensive review.

On the basis of the properties summarized above, it may not be a surprise that FeRh has received attention in view of possible applications. Indeed, FeRh has been investigated as magnetocaloric material for efficient magnetic refrigeration [10] and, more intensively, for spintronic applications. For instance, Thiele et al. [11] demonstrated that FeRh can be of interest in thermal-assisted magnetic data storage, and soon after Thiele et al. [12] and Ju et al. [13] demonstrated that the AFM--FM phase transformation can be driven by ultrashort laser pulse. More recently, within the framework of the emerging field of AFM spintronics, FeRh has gained a renewed attention by the demonstration that the spin-orbit coupling and the associated anisotropic magnetoresistance (AMR) in the AFM phase, can be used to create a memory resistor robustly storing hidden magnetic information in the AFM state [14].

For spintronic applications, the magnetostructural properties of FeRh can be most easily exploited by taking advantage of the sensitivity of the properties of FeRh to cell volume modifications, as anticipated by the experiments performed under hydrostatic pressure mentioned above. Following this approach, voltage-control of magnetic and electric properties of FeRh have been attempted in recent years, with demonstration of a large modulation of magnetization and resistivity under electric field [15,16]. Electric control of magnetism appears to be a promising avenue for energy efficient magnetic data storage and processing, as revised recently by Taniyama [17] and Hu et al. [18] and by the excellent review by Meisenheimer et al. [19].

Here, we aim at reviewing the state-of-the art of voltage control of magnetic and electric properties of thin films of FeRh alloys. To be self-contained, the review starts by revising the knowledge on strain effects caused by epitaxy or clamping on substrates and by capping layers, on the properties of FeRh films (Section 2). In Section 3, we focus ubiquitous observation of AFM and FM phase coexistence at T < T*, which is particularly relevant in thin films. In Section 4, we review the nucleation and growth of AFM/FM domains across the AFM-FM transition and its impact on properties. Armed with this knowledge, in Section 5 we focus on the state-of-the-art of FeRh films grown on piezoelectric substrates and in Section 6 we address the issues of the dynamics of AFM-FM phase transformation and memory effects in FeRh-based devices. We conclude in Section 7 with a summary and some outlooks.

## 2. Epitaxial strain in FeRh thin films

Early attempts to grow thin films of FeRh revealed that the magnetic transition was incomplete and displayed a large thermal hysteresis [20], which contrasts to the sharp transition observed in bulk materials. Ohtani et al [21] prepared FeRh films, which turned out to be under tensile stress. They observed that releasing the tensile stress by peeling, produced an increase of T* of about 50 K, thus suggesting that the pre-existing tensile stress had favored the FM state; this observation could be consistent with hydrostatic measurement [4]. However, the polycrystalline nature of films of Ohtani et al. precluded a deeper understanding of the role of strain on thin films.

It was not until 2005 that the preparation of FeRh epitaxial films was reported [22]. Maat et al. grew films about 110 nm thick, on MgO(001) and $Al_2O_3$(0001) substrates, having (001) and (111) textures respectively. The in-plane and out of-plane cell parameters of FeRh were extracted. The films on MgO were found to be slightly expanded out-of-plane (c) and compressed in-plane (a) (tetragonal c/a > 1 distortion), while the films on *c*-sapphire ($Al_2O_3$) were compressed out-of-plane and expanded in plane due to the larger in-plane cell parameters of $Al_2O_3$(0001) compared to MgO(001). As the cell parameters of the FM phase of FeRh are larger than those of the AFM phase, it can be expected that the films on c-sapphire have a lower T* (FM phase is more stable) than on MgO. Indeed, this was the experimental observation: T*($Al_2O_3$) ≈ 350 K and T*(MgO) ≈ 375 K. However, it is worth noticing that the unit cell volume of FeRh was found to be about the same on both substrates, suggesting that unit cell volume is not the key parameter dictating T*. It was observed that the AFM to FM transition (and vice versa) was largely hysteretic as expected for a $1^{st}$ order transition and very broad (about 50 K) probably due to crystalline imperfections and chemical disorder. These behaviors has been ubiquitously seen in most of subsequent reports on FeRh thin films.

Xie et al. [23] reported the growth of epitaxial FeRh films (≈ 80 nm thick) on (001)$SrTiO_3$ (STO), (001)MgO and (001)$LaAlO_3$ (LAO) substrates. Of relevance is that the unit cell of STO (3.905 Å), MgO (4.216 Å) and LAO (3.792 Å) are different from that of the cubic FeRh (2.985 Å, in the AFM phase, 2.995 Å in the FM phase) and the corresponding structural mismatch may impose an epitaxial strain of different sign. It was found that FeRh grew epitaxially on STO and MgO with a 45° in-plane rotation: (FeRh(001)<110>//STO(001)<100> and FeRh(001)<110>//MgO(001)<100>. The corresponding structural mismatches are: f = -7.8 % (STO) and f = -0.53 % (MgO). The small mismatch for MgO can be accommodated by a compressive strain whereas for the case of STO a coherent compressive stress cannot be stabilized across the whole film, but some structural relaxation should occur. In the case of LAO the optimal structural mismatch occurs for cube-on-cube growth, leading to a tensile (f = +1.2 % mismatch). It was found that these films (upon warming) have T* = 405 K (FeRh/STO, f = -7.8 %), T* = 393 K (FeRh/MgO, f = -0.53 %) and T* = 360 K (FeRh/LAO, f = +1.2 %). These experimental results indicate that the compressive and tensile strain can lead to an increase and decrease of the AFM to FM transition temperature, respectively. Unfortunately, an accurate determination of in-plane and out-of-plane cell parameters was not reported and thus the agreement of experimental results with expectations remained rather speculative.

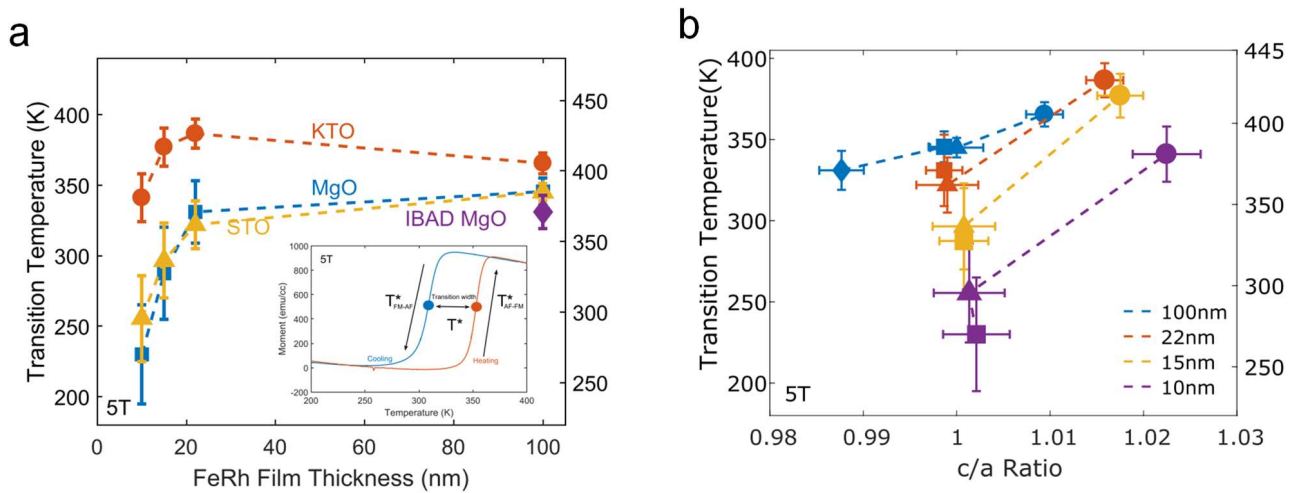

**Figure 1.** (a) Transition temperature (T*) as a function of film thickness of FeRh films grown on different substrates as indicated, recorded under a 5 T field (left axis) and adjusted to zero field (right axis). Circles, squares, triangles, and diamond correspond to FeRh films grown on KTO, MgO, STO, and IBAD MgO, respectively as indicated. Colored symbols and connecting dashed lines help to visualize the data for different substrates. The inset illustrates the magnetization as a function of temperature for a 22 nm FeRh film grown on MgO. (b) The transition temperatures as a function of the c/a ratio for films of various thicknesses. Dashed lines in b) connect sample of similar thickness grown on different substrates. Tensile strain corresponds to c/a<1 and compressive to c/a>1. See Ref. [24] for details. Reproduced with permission from ref. [24].

Ceballos et al. [24] explored the dependence of T* on epitaxial strain and thickness by growing films of different thicknesses (10, 15,22 and 100 nm) on cubic substrates having different lattice parameters: (001)STO, (001)MgO and (001)KTaO$_3$ (KTO), all of them imposing compressive strain on FeRh. The cell parameter of KTO (= 3.989 Å) is intermediate between those of STO and MgO. Tensile strain effect were explored using (001)MgO layers prepared by high energy ion-beam assisted (001)MgO (IBAD-MgO). It was found [Figure 1(a)] that T* decreases with decreasing film thickness below 22 nm for all substrates and T* was found to depend on the strain acting on the films. The inset shows the magnetization vs temperature M(T) data for a 22 nm film grown on MgO and indicates the transition temperatures upon cooling and heating in a 5 T field. It was observed that different strain states (imposed by the substrate) lead to different tetragonality ratios (c/a) and that the (c/a) ratio is the parameter ruling T*. Inspection of Figure 1(b) allows to conclude that when increasing c/a, T* also increases, signaling the stabilization the AFM phase. Kumar et al [25] obtained similar results.

To avoid surface oxidation, FeRh films are quite commonly capped with different layers, either insulating (for instance MgO or AlO$_x$) or metallic (Al, Au, Cr, W, etc) that may impact T* and promote the occurrence of interfacial FM [31]. We focus here on the effects of strain on FeRh due to the combined action of the substrate and capping layer. Loving et al [26] reported a systematic study of capping effects on FeRh films (30 and 50 nm thick) grown on (001)MgO and (0001)Al$_2$O$_3$. Capping layers of AlO$_x$, Al, Au, Cr and W of different thicknesses (0.15- 8 nm) were used and the c/a ratio was derived from X-ray diffraction experiments. Interestingly, it was found that, for a given substrate, the capping layer appear to modify c/a ratio thus impacting T*. Figure 2(a) shows an illustrative example and Figure 2(b) collects a summary of T*(c/a) data from literature (see ref. [31] for details). It is worth noticing that most samples in Figure 2(b) have different tetragonality but the volume of the unit cell remains constant [Figure 2(c)]. The message that emerges is that the different strain, more precisely the c/a, resulting from the different capping materials determine T* in a similar manner than substrate-induced epitaxial strain.

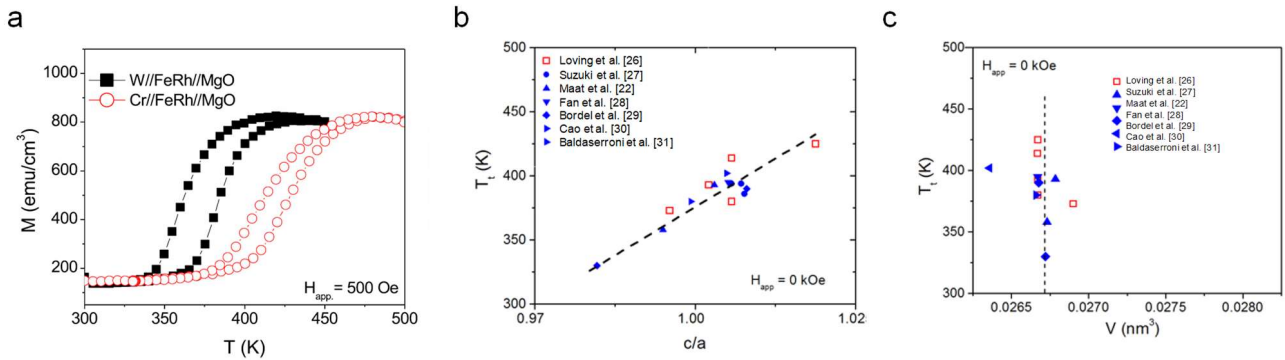

**Figure 2.** (a) Magnetization data collected as a function of temperature ($H_{app}$ = 500 Oe) for the Cr/FeRh/MgO and W/FeRh/MgO films. (b,c) Dependence of the magnetostructural transformation temperature ($T_t$) of thin film and bulk FeRh systems as a function of film lattice distortion (*c/a*) and unit cell volume (V), respectively. Data from the panels (b) and (c) were extracted in ref. [26] from refs. [22,26,27,28,29,30,31]. Reproduced with permission and adapted from ref. [26]

### 3. Phase coexistence in FeRh thin films

It was early recognized that in FeRh thin films, the transition from the high temperature FM phase to the AFM low temperature phase may not be complete. In spite of the transition being well visible in magnetization *vs* temperature M(T) curves, incomplete FM-to-AFM transition is evidenced by the presence of a residual magnetization at T < T*. Early in 1999, van Driel et al. [32] reported a comprehensive study of the compositional dependence of the magnetic properties of $Fe_xRh_{1-x}$ films. It was made clear that magnetic transitions in thin films are broader than in bulk materials of similar compositions, and suggested that this could be due to local changes of $Fe_xRh_{1-x}$ composition. In fact, for Rh-rich compositions the magnetization was found to decrease and T* lowers [32]. Suzuki et al. [27] had already noticed that the residual magnetization in the AFM phase, was particularly noticeable when reducing film thickness, thus suggesting that the FM phase is more stable when narrowing film thickness. Although this result could be in agreement with the predictions made by Lounis [33], who argued that upon decreasing thickness the FM state is the ground state of FeRh, the observation that although a FM signal is observed at low temperature, still the transition FM to AFM was well visible, casting doubts on the relevance of the theoretical prediction on the observed effects.

These data may also suggest that the interfaces, either with the substrates or the top one, either free or *ad-hoc* capped, may display some FM signatures, being naturally more relevant when decreasing film thickness. Ding et al. [34] had previously prepared FeRh/MgO films capped with MgO or Au. Magnetization data indicated an abrupt transition from AFM to FM states, occurring between 340 to 360 K (upon heating). As in Ref. [27], an unexpected small FM component was observed at room temperature. Ding et al. used X-ray magnetic circular dichroism (XMCD) either in total-electron yield (TEY) or indirect-transmission (IT) modes to explore the magnetic signal at room temperature. Of relevance is that IT is bulk sensitive but TEY is rather surface sensitive (4-10 nm). It was found that a FM signal is well visible in TEY but not in IT, thus suggesting that the FM signal, that amounted about 5% of the total saturation magnetization of FeRh, was originated at the surface in contact with the capping layer. Interestingly, it was found that the FM component was one order of magnitude larger for MgO capping than for Au capping. This intriguing difference, which may have far reaching implications for applications and fundamental science, remains to be fully understood.

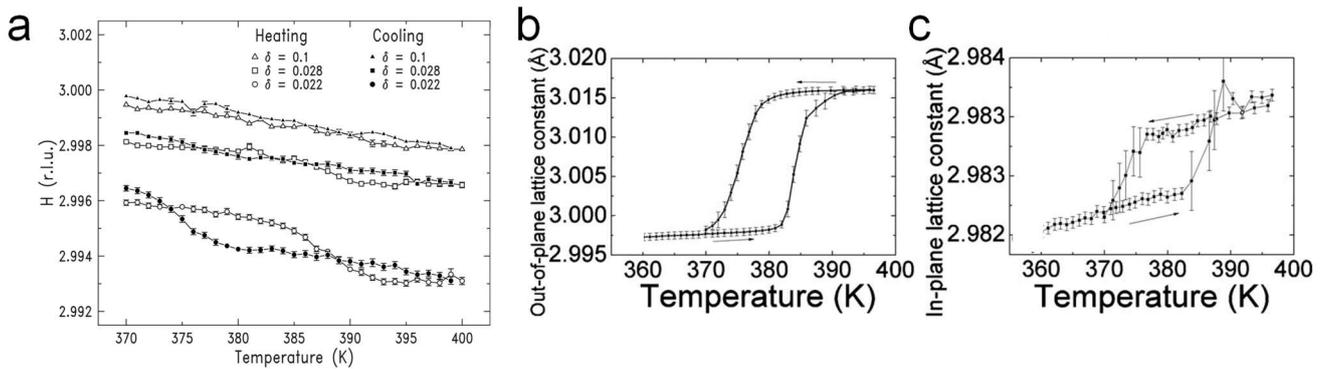

**Figure 3**. (a)Temperature dependence of the H position for the in-plane (3 0 δ) refection of the FeRh film at various δ = 0.022 (circles), 0.028 (squares) and 0.1 (triangles), which correspond to scattering depths Λ=1.8, 2.1, and 50 nm, respectively. Reproduced with permission from ref. [35]. (b,c) Out-of-plane lattice constant and in-plane lattice constant across the magnetic transition Reproduced with permission from ref. [28].

In any event, data suggested that FM regions exist at the film surface, may be exacerbated by the capping layer. To get insight into the origin of this surface effect, Kim et al. [35] performed high-resolution grazing-incidence X-ray diffraction measurements on uncapped FeRh/MgO films grown by molecular bean epitaxy (MBE). In Figure 3(a) (reproduced from [35]), the temperature dependence of the position of the (30δ) reflection of FeRh, recorded along the surface truncation rod of the (300) reflection at various depth (given by δ), is shown. It is clear that for the most bulk sensitive measurements (δ = 0.1 and 0.028), only the conventional thermal expansion of unit cell, clamped at the substrate, is visible. In contrast, for most surface sensitive data (δ = 0.022), a clear change of the in-plane cell parameter is observed at the magnetic transition, which reflect that the surface of the FeRh is somewhat relaxed. It is at the free surface where the FM transition starts probably due to the strain relaxation.

Polarized neutron reflection experiments (PNR) were used to explore the magnetism of the top and bottom interfaces in FeRh/MgO(50 nm) films capped with MgO [28]. The temperature dependence of the out-of-plane and in-plane cell parameters are shown in Figure 3(b,c). We notice in Figure 3(b,c) that the cell expansion occurring when entering into the FM phase, is more pronounced along the out-of-plane than along the in-plane direction, due to the epitaxial clamping of FeRh to the substrate. However, at variance with data in Figure 3(a) the change of phase is well visible in the in-plane cell parameters of Figure 3(c), although it is mentioned in ref. [28] that the samples are the same. In any event, the important point is that the film, at 300 K, well into the AFM region, has a magnetization of about 3.75% of the fully saturated FM phase, and the question is where this FM phase reside. From the PNR data it was concluded that a region of about 7-8 nm at the bottom FeRh/MgO interface is robustly FM at room-temperature without signs of interdifusion, whereas a weaker FM signal arises from the FeRh-MgO capping interface, which appears to be associated to chemical interfusion (≈10 nm) and to a gradient of Fe:Rh composition [32].

Zhou et al. [36] imaged *in-situ* the surface and the bulk magnetic signal of an *uncapped* FeRh/MgO(001) films by scanning electron microscopy with polarization analysis (SEMPA) and depth-sensitive magneto-optical Kerr effect, respectively, as a function of temperature while crossing the FM-to-AFM phase transition. It was observed that while the FM-to-AFM at the bulk of the film was complete, FM was persistent at the surface at T<<T*. Thus, it was concluded that the existence of a FM surface layer is an is an intrinsic property of the FeRh(001) surface, rather than an artifact induced by capping layers or native surface oxide layers.

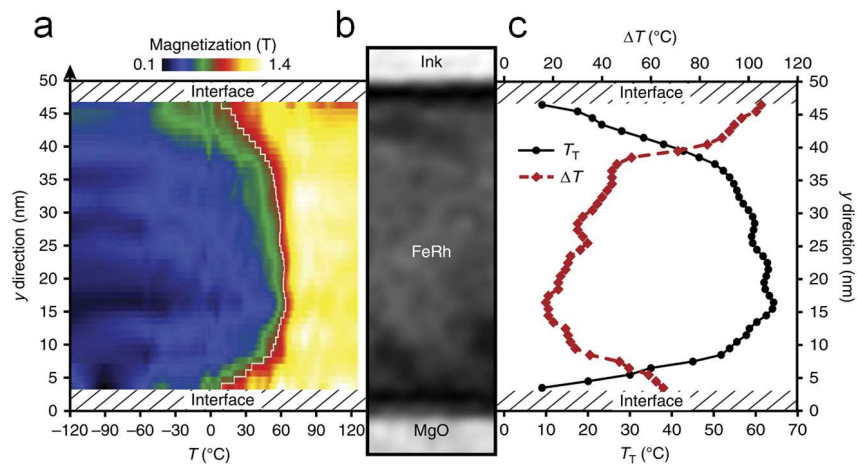

**Figure 4.** Evolution of the magnetization in the growth direction. (a) Map of the magnetization as a function of the temperature and the position in the layer depth using the heating temperature series. The color scale corresponds to the magnitude of the magnetization and the transition temperature is displayed by the white profile corresponding to magnetization of 0.75 T. (b) Amplitude image of the part of the FeRh layer used for the calculation. (c) Profiles of the transition temperature T* and the transition width ΔT as a function of the position within the FeRh layer. Reproduced with permission from ref. [37].

Gatel et al. [37] recently revisited the long standing question of the spatial distribution of magnetic patches in FeRh films. They used electron holography), a technique that allows quantitative mapping of the local magnetization and permits extraction of magnetic information across the entire film thickness with unrivaled spatial resolution. The magnetic profile across the film was determined as a function of temperature (Figure 4(a)) in a certain region (Figure 4(b)). Data were used to determine the depth-dependence of T* [Figure 4(c)]. As commonly observed, Gatel et al. also identified FM regions, existing at T<<T*. More importantly, they notice that upon heating, the FM magnetic signal at both interfaces, starts increasing at T<T*. This implies that T* is lower at interfaces than at the center of the film, and thus FM regions nucleate first at both interfaces.

### 4. Nucleation and growth of AFM/FM domains

Nucleation and growth of FM phase has been studied mostly by using XMCD and photoemission spectroscopy (PEEM) (XMCD-PEEM). Baldasseroni et al. [38] first reported images showing the evolution of the FM (FM) domains across the temperature-driven AFM to FM phase transition in uncapped and Al-capped epitaxial FeRh thin films. The coexistence of the AF and FM phases was evidenced across the broad transition and the different stages of nucleation, growth, and coalescence were directly imaged. It turned out that the FM phase nucleates into single-domain islands and the width of the transition of an individual FM nucleus is sharper than that of the transition in a macroscopic average. Figure 5(a-d) shows representative images of both films near room temperature (AFM state) and above 400K (FM state) taken at the Fe $L_3$ edge. While the high-temperature image of each film [Figures 5(b) and (d)] shows a similar complex pattern of micron size FM domains with strong magnetic asymmetry (i.e. magnetic circular dichroism) ranging from -0.15 to 0.15 (the same order of magnitude asymmetry than that of pure Fe at the Fe $L_3$ edge), a striking difference can be observed between the two room-temperature images shown in Figs. 4(a) and 4(c). The uncapped sample shows no magnetic contrast in the AFM phase, whereas the capped sample clearly shows FM domains, albeit with reduced contrast compared to the full FM images. Since the oxide layer of the uncapped sample is thinner than the probing depth, the vanishing magnetic contrast of the uncapped sample is attributed to the absence of FM regions within the AFM phase of FeRh, while the domains of the capped sample reveal a stable FM phase at the interface with the Al capping layer. In fact, interfacial ferromagnetism at room temperature have been previously observed [28,34], as described above. The stabilization of the FM phase at the interface is claimed to result from a combination of Fe deficiency and strain effects due to the cap, although the fine details remain to be elucidated. On the other hand, these data suggested that a thin native oxide layer does not promote surface FM or it is too weak to be detectable. The coexistence of FM and AFM phases inherent to a first-order phase transition was observed in the early stages of the transition and small FM nuclei were seen to form flux closure patterns. If this observation could be taken as representative of the nucleation and growth of all FM nuclei, it would imply that the net magnetization of FM would vanish and thus the sample, although being FM will display a zero magnetization in zero field. It will be shown latter that radically different results have been obtained by other authors.

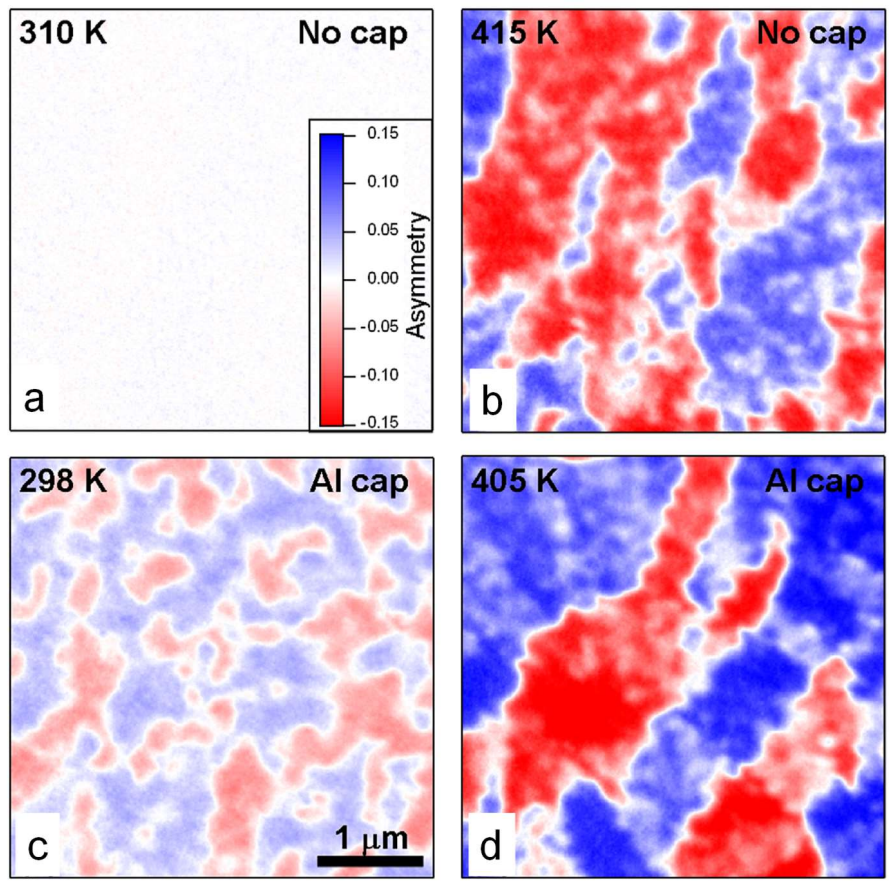

**Figure 5** Effect of capping layer and temperature on FM domains in FeRh thin film. XMCD-PEEM asymmetry images (difference between right and left polarization images divided by their sum, in zero applied magnetic field) of FeRh thin films showing FM domains in the AF and FM states (temperature of image shown in upper left corner) for the film without capping layer [No cap (a) and (b)] and the film capped with 2.5 nm of Al [Al cap (c) and (d)]. FM domains with a positive (negative) projection of the magnetization onto the x-ray beam direction are shown in different intensities of blue (red). The asymmetry color scale used for the 4 images is shown in (a). Reproduced with permission from ref. [38].

Nucleation, growth and coalescence of FM nuclei were identified in the phase transformation from AFM to FM. By using measurements of out-of-plane lattice constants in FeRh/MgO, the coexistence of lattice-expanded (FM) and lattice contracted (AFM) phases in spatially distinct regions of the samples was revealed [35,39]. Interestingly, the phase separation was found to be more pronounced during cooling than heating, and it was observed that the FM phase can be undercooled several Kelvins but there is no equivalent superheating of the AFM domains. This behavior was understood as being reminiscent of the solid-liquid phase transitions, where supercooling in liquid-to-solid transition is commonly seen, because solid nuclei within a liquid matrix must exceed some critical size before being kinetically stable, but the reverse is not true as liquid droplets are more easily present at the surface of the solids when approaching the melting temperature and they grow rapidly [39]. However, identification of the microscopic mechanisms of the melting/freezing process remained unsolved. The asymmetric *melting* and *freezing* processes across the AFM-FM transition was also imaged by PEEM. Baldasseroni [40] showed that the formation of AFM domains upon cooling is dominated by the nucleation process at defects, with little subsequent growth, resulting in a non-random distribution of small AFM domains. Indeed, repetitive cooling process indicate nucleation of AFM nuclei at fixed positions in the film, reflecting their pinning within the FeRh film. In contrast, upon heating, there is a heterogeneous nucleation of FM domains at different sites followed by a subsequent growth in size. The asymmetry of the AFM-FM transition is also visible in the resistivity changes across the transition and it was found to be exacerbated in mesoscale strips (about 1 μm wide) patterned on 25nm FeRh/MgO films [41]; it was observed that

I. Fina and J. Fontcuberta

resistivity measurements displayed pronounced supercooling and avalanche-like abrupt transition from FM to AFM whereas the reverse transition was found to be continuous and gradual. It was argued that the different dynamics of nuclei upon AFM to FM transition and vice versa can be understood as mainly due to the different strength of the magnetic correlations within the FM and AFM regions. Indeed, FM correlations are generally robust to local disorder and even persist in granular or amorphous films, whereas AFM systems typically show shorter-range correlations and limited by crystalline correlation lengths. In this context, resistivity steps in the warming process are understood as being originated by the individual AFM regions undergoing a first order transition to the FM state. The robustness of the FM state under cooling is responsible for the supercooling until the AFM nucleates and by breaking the FM correlations in the meso-strips, produces a sharp transition. In short, the asymmetry of the AFM-FM transitions seems to be governed by the nature of the magnetic interactions in the corresponding matrices (FM or AFM).

The clearest evidence of the nucleation and growth process of FM/AFM nuclei were obtained by Keavney et al. [42] reporting a study of FeRh(20 nm)/MgO films by combining nano X-ray diffraction data with a spatial resolution of ≈ 30 nm, and XMCD-PEEM with spatial resolution of ≈ 100 nm. Figure 6(a) below displays nano-XRD data images of the film taken in a heating process. In these images it can be appreciated that FM domains nucleate within the AFM matrix upon heating. Interestingly, some domains, see for instance the one indicated by an arrow at 370 K, and similarly the AFM domain within the FM matrix (see arrow at 376 K), evolve slowly with temperature, suggesting that nucleation is driven by defects in the lattice and the region surrounding the defects has a lower transition temperature probably due to localized strain. The subsequent growth of each island is apparently restricted by the surrounding material, which implies that material not neighboring a defect has a higher transition temperature. This kinetic arrest of the domain growth is attributed to the competition of energetically favorable strain states [42], and again points to the relevant role of strain on the AFM-FM transition.

The role of defects as nucleation centers of the FM/AFM domains can be well appreciated in Figure 6(b) were the nano-XRD and XMCD-PEEM data of a given region in the film are compared after different thermal cycling processes [42]. Data clearly show that structural and magnetic domains appear at the same position upon cycling.

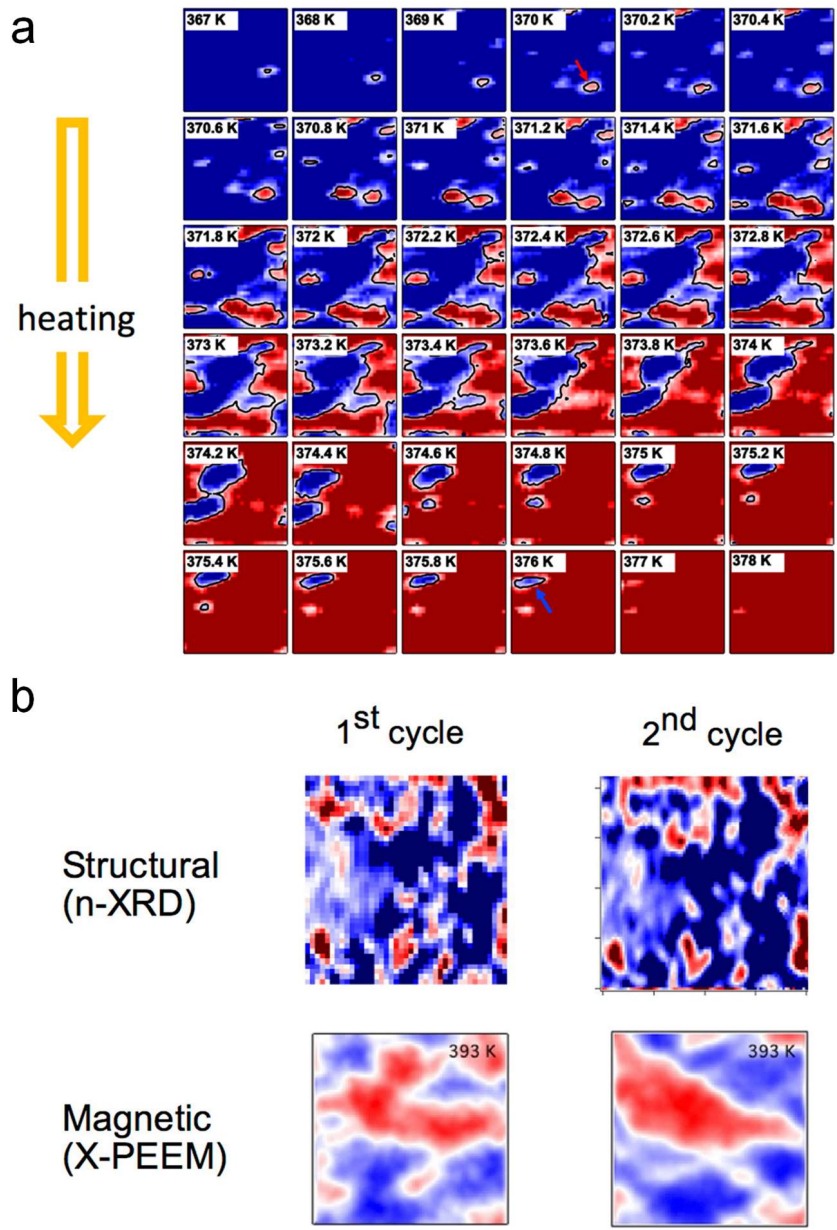

**Figure 6** (a) Temperature dependent nano-XRD data taken at the FeRh(002) reflection while heating across the transition dependent taken with a 30 × 30 nm$^2$ focused spot. Red (blue) indicates the low (high) temperature phase of FeRh. (b) Nano-XRD images taken at the midpoint of the structural transition (upper) and XMCD-PEEM images taken in the high temperature phase (lower) for two consecutive heating and cooling cycles. Note that the structural and magnetic images show two different contrast mechanisms. All images are 4 × 4 μm. Reproduced with permission from ref. [42].

## 5. FeRh films on piezoelectric substrates

Having stablished that the AFM-FM transition occurring at T* is extremely sensitive to epitaxial strain, with compressive strain producing an increase of T* and tensile strain an lowering of T*, a natural consequence could be to suspect that if FeRh films were grown on materials displaying structural transitions, the change of lattice parameters of the substrate may have visible consequences on the magnetic and transport properties of FeRh films. This was the approach of Suzuki et al. [43] who grew Ga-FeRh films on single crystalline ferroelectric BaTiO$_3$(001) substrate. BaTiO$_3$ (BTO)undergoes three structural phase transitions from the paraelectric cubic (C) to the ferroelectric tetragonal (T) phase at 393 K, T to the orthorhombic (O) phase at 278 K and O to rhombohedral (R) phase at 183 K. To promote an enhancement of the possible effects of strain changes on the properties of the magnetic alloy, Ga-substituted FeRh was used because it had been previously reported that Ga substitution lowers T* down to about 270 K, thus approaching it to the T-O structural transition of BTO. Consistent with expectations, it was found that the temperature-dependent magnetization M(T) of the film displayed distinctive changes at the phase-transition temperatures of the substrate. More precisely, M(T) curves measured upon cooling and using a small magnetic field (500 Oe), exhibited noticeable changes at 285 K and 190 K, coinciding with the structural transitions of BTO. Interestingly at the T-O transition the magnetization was found to increase; however, at the O-R transition it decreases [Figure 7(a)], which contrast with the monotonic variation of magnetization observed in the field-dependent magnetization loops M(H) [Figure 7(b)]. This different behavior arises because a small field, smaller than the anisotropy field, was used in the measurements in Figure 7(a). Moreover, in Figure 7(b) it can be appreciated that the saturation magnetization measured in the tetragonal phase of BTO (290 K) is larger than that recorded in the orthorhombic phase (285 K). This trend is opposite to that observed in the M(T) data recorded at low field [Figure 7(a)]. The different behavior can be understood by noticing the different shape of the low-field region of the magnetization loops recorded at 290 K and 295 K.

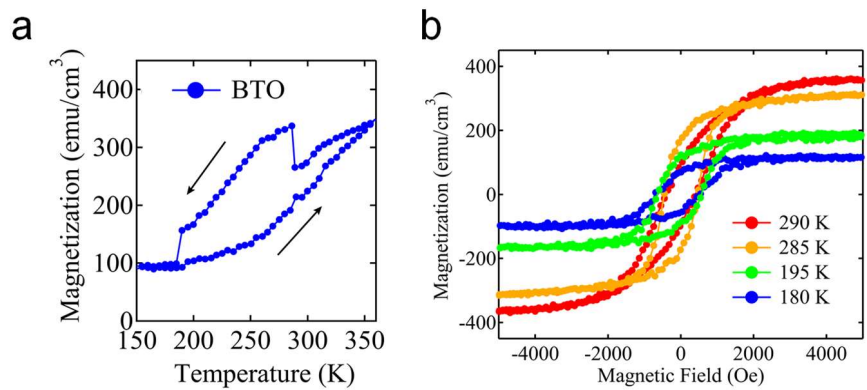

**Figure 7** (a) Temperature dependence of magnetization of (a) Ga-FeRh/BTO measured in a magnetic field of 500 Oe applied in-plane along [110] of Ga-FeRh. (b) Field dependence of magnetization of Ga-FeRh/BTO at various temperatures. Reproduced with permission from ref. [43].

Results were interpreted by arguing that at T-O there is a change of the magnetic anisotropy axis but not any strain-induced FM to AFM phase transition. In contrast, at the O-R transition, the compressive stress imposed by the R phase on Ga-FeRh dictates AFM as the new ground state and thus the magnetization of the sample lowers. These results illustrate the strain not only shifts T* but also changes the anisotropy of the magnetic phase. Chen et al. [44] and Bennett et al. [45] reported similar results using FeRh rather than Ga-FeRh. In addition, Chen et al. performed temperature-dependent X-ray diffraction experiments and showed that both, the change of lattice parameters of each phase (T,O,R) together with the domain reconstruction at the different phase transitions, in overall, produce a compressive strain that, upon cooling stabilizes the AFM state.

Thus, the obvious next step was to attempt to exploit the piezoresponse of piezoelectric substrate to allow *in-operando* tailoring of the magnetic response. Indeed, voltage-control of magnetic phase transition in FM shape memory alloy had early been demonstrated by Chen et al. [46] by epoxy bonding the alloy on a piezoelectric single crystal. Cherifi et al. [16] succeeded to grew FeRh on a $BaTiO_3$(001) single crystal and to demonstrate a voltage-control of the magnetic properties of FeRh films. The M(T) curves measured under applied voltage show that T* is pushed to higher temperatures. Consequently, at a given temperature, a giant change of magnetization (ΔM) is observed, for a modest electric field. Consistently, isothermal magnetization data (385 K) recorded under electric field lead to a significant change of magnetization [Figure 8(a)]. Here, the crucial point is that T* have been shifted by the applied E, changing the relative abundance of AFM/FM domains in the film and therefore the magnetization. The role of the electric field, applied along de c-axis, is multiple. First, it transforms the initial multidomain BTO single crystal, containing a- and c-domains in its virgin state, into a single domain c-type sample. The change of domain populations from (a,c) to (c) implies a contraction of the average in-plane parameters of BTO and consequently the FeRh film also contracts in-plane. According to the results derived from the role of epitaxial strain discussed above, an in-plane compression should favor an increase of T*, reflecting the stabilization of the AFM phase and thus a reduction of the sample magnetization. The hysteretic domain reconstruction occurring upon pooling the BTO shall produce a butterfly-like loop in the magnetization of the FeRh film if strain coupling is present, as observed in Figure 8(a). The evolution of the out-of-plane cell parameters of FeRh [see inset in Figure 8(a)] confirms this view. Latter, Phillips et al. [47] reported XMCD-PEEM images of the FeRh/BTO sample, performed under V-bias, which confirmed the major role played by the switching and reordering of ferroelastic (a,c) domains on the FM to AFM transition. In passing, we strength that the typical sizes of these domains are of tens to hundreds μm, much larger than the FM domains in FeRh. As lengthy reviewed in an earlier section, the unit cell contraction of FeRh brings an associated stabilization of the AFM phase. A second effect, is that under E-field the piezoresponse of a fully polarized BTO crystal should produce an additional in-plane compression, with a reversible expansion/compression under bias. Unfortunately, this regime was not explored in [16]. A third possible contribution, refers to a genuine electric-field effect on the carrier density of FeRh. Indeed, the magnetic properties of FeRh are known to be extremely sensitive of carrier concentration [48]. Cherifi et al. [16] attributed the slight antisymmetric variation of magnetization for V > 0 and V < 0, observable in data of Figure 8(a), to this electric field effect. At first sight, electric field effects in metals are not expected to be relevant as field is screened within a distance from the surface given by the Thomas-Fermi screening length ($\lambda_{TF} \approx 0.3$ nm in FeRh). Cherifi et al. argued, however, that in magnetic materials, effects can be observed at distances set by the exchange length interaction which is much larger than $\lambda_{TF}$. This view was not shared by Liu et al. [15] who concluded that the electrostatic effect on charge modulation in FeRh and so on its electrical and magnetic properties, should be negligible. Although data in Figure 8(a)(inset) indicates a c-axis expansion under V-bias and a concomitant change of magnetization, it remains to be seen if the unit cell volume varies or not under the strain induced by the switching of the ferroelastic domains, and if this possible change has a role on the AFM-FM transition. Liu et al. [15] addressed this issue by performing electric resistivity measurements of FeRh/BTO films to determine T* while biasing the sample with a suitable electric field. They complemented this study with the determination of the in-plane and out-of-plane cell parameters of FeRh under V-bias. Figure 8(b) shows the dramatic changes of resistivity (ρ) occurring at T* and a giant change of resistivity Δρ/ρ by about 22% induced by 2 kV/cm. This is the electric counterpart of the magnetic changes shown in Figure 8(a). Interestingly, the out-of-plane cell parameter was found to be expanded under V-bias, while the in-plane parameter contracted due to BTO domain switching. Consequently, the tetragonality of the unit cell of FeRh increased by about a 0.72%. However, the unit cell volume only contracted by about -0.17%, which is much smaller than the changes observed at the AFM-FM transition (≈ -1 %). Therefore, it was concluded that tetragonality rather than unit cell volume compression is the relevant parameter for the magnetic transition, in agreement with the results obtained growing FeRh films under different substrates imposing different strain summarized in Section 2.

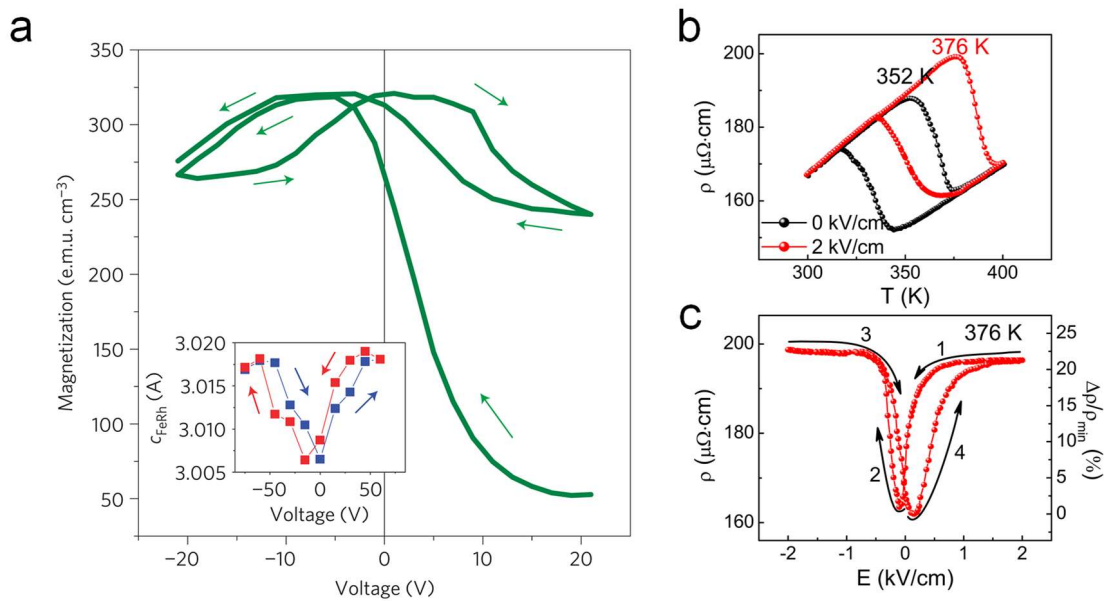

**Figure 8** a) Voltage dependence of the magnetization and the structural parameters. **a**, Variation of the magnetization with the applied voltage at 385 K, after warming to this temperature in a voltage of 21 V. The inset shows the voltage dependence of the out-of-plane parameter of FeRh at 390 K. Reproduced with permission from ref. [16]. b) ρ-T curves with and without a gate E between 300 and 400 K. (c) Resistivity enhancement Δρ(E) under E=2 kV/cm as a function of T during warming and cooling. The ratio of Δρ(E) over the resistivity difference between the AFM and FM phases Δρ(AFM-FM)∼35 μΩ·cm is marked on the right axis. (d) ρ versus E measured at 376 K by scanning E from 2→−2→2 kV/cm. The electroresistance (right axis) is normalized to the resistivity minimum. Reproduced with permission from ref. [15].

Similar experiments have been reported using FeRh films grown on single crystalline PMN-PT(001) substrates. PMN-PT is a ferroelectric relaxor of composition $0.72PbMg_{1/3}Nb_{2/3}O_3$-$0.28PbTiO_3$. Lee at al. [49] found that the electric resistivity of the FeRh film displays large variations when a biasing voltage is applied across the structure. It was observed that Δρ/ρ changes by about 8% under 6.7 kV/cm. This variation of resistance is significantly smaller than that recorded in FeRh/BTO [15].

Still using FeRh/PMN-PT heterostructures, Xie et al. [50] was able to demonstrate reproducible and symmetric changes of the coercive field of FeRh, displaying a butterfly loop. They concluded that the magnetic anisotropy was not changed by the piezo strain and thus they concluded that the change of coercivity was mainly due to the strain-triggered shift of T*.

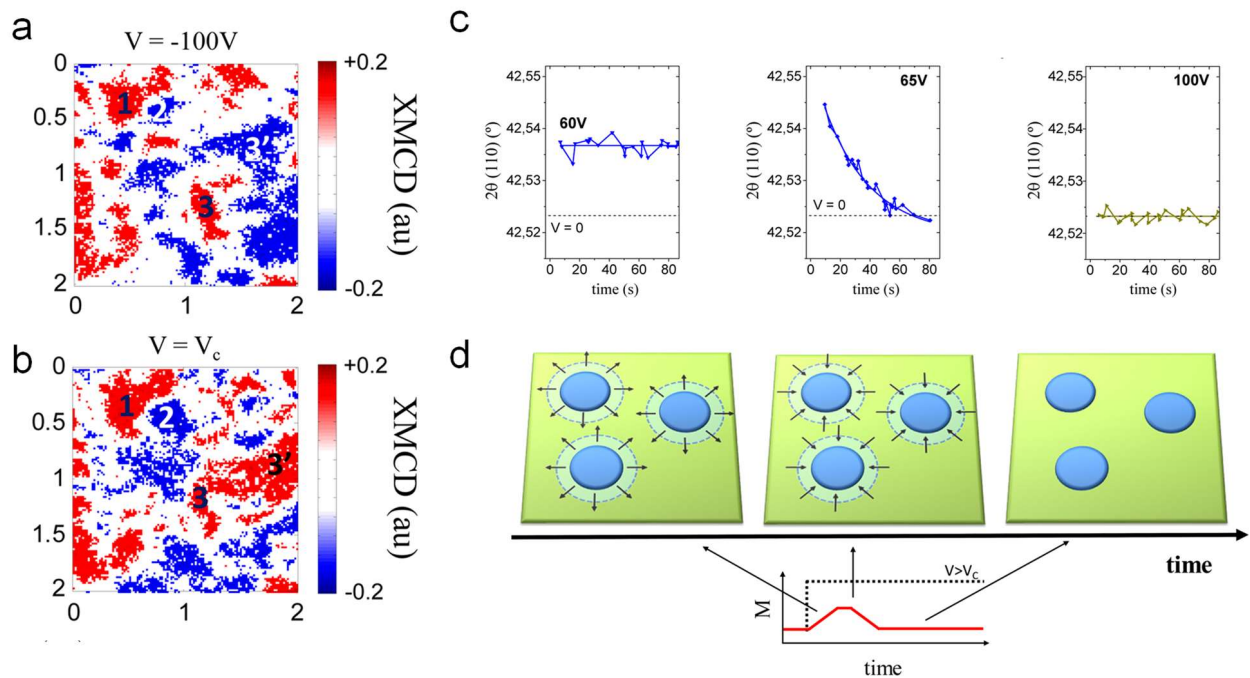

**Figure 9.** a,b) XMCD–PEEM image collected in the very same region at V = −100 V and V = $V_C$, respectively. The false color scale corresponds to the projection of the magnetization onto the incident X-ray beam direction (horizontal from the left). Domains with magnetization parallel and antiparallel to the X-ray incidence have opposite contrast (blue and red regions). Perpendicular domains or domains with zero magnetic moment appear white. Images correspond to 2 μm lateral size. **(c)** Time-dependent FeRh (110) peak position 2θ(110) as a function of the delay time after switching on the bias V = 60, 65 and 100 V as indicated. V. A prepoling field V = -100 was used in all cases. Lines through data are a guide for the eye. **(d)** Expansion compression of the FM domains during the transient magnetoelectric response for voltages just above $V_C$ and final recovery of the initial state. Reproduced with permission from ref. [15]from ref. [54].

## 6. Dynamics of AFM-FM phase transformation and memory effects

As first-order transitions are driven by nucleation and growth of domains and these processes need a certain time to take place, a time-dependent response of the should be observable. For instance, in a magnetic field or a temperature-induced transition from AFM to FM, if the magnetic-field or heating/cooling rate are too high, the material simply cannot follow and, consequently, a broadening of the hysteresis across the transition is commonly observed which depends on the rate of variation of the corresponding stimulus. The phase coexistence of AFM and FM nuclei in some region of the space implies that the system is in a metastable state and thus a time dependence of the magnetization, should occur. Indeed, the metastability has been reported in many materials displaying first order AFM-FM transitions such as magnetic shape-memory alloys [51] and the Laves phase CeFe$_{2-x}$Ru$_x$ [52]. FeRh is not an exception [Feng13-36] [53]. The dynamics of the transformation needs to be considered, especially for potential applications.

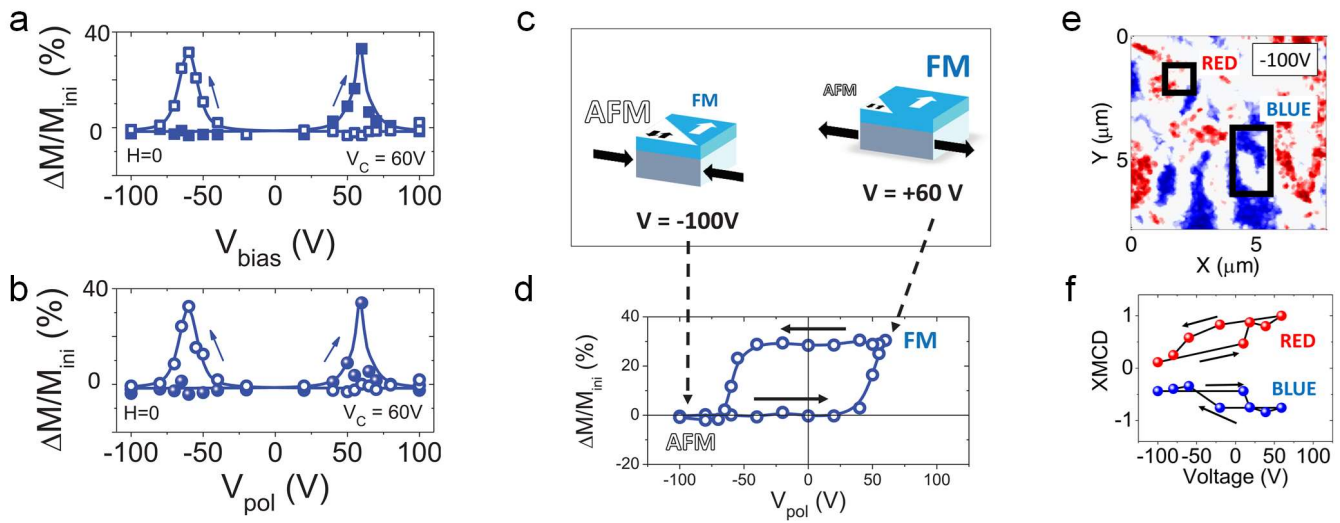

**Figure 10.** (a) Relative increase of magnetization of with respect to the magnetic moment in the initial state [$\Delta M/M_{ini}$ = M–M$_{ini}$)/M$_{ini}$] depending on bias voltage (V$_{bias}$) applied during the measurement. (b) Relative increase of magnetization with respect to the magnetic moment in the initial state at electric remanence (0 V) depending on previously applied bias voltage (V$_{pol}$). (c) Sketch of the two remnant magnetization states obtained for the two different strain states. The increase/decrease in the substrate in-plane strain induced by the application of the external voltage (poled or V$_{bias}$ = V$_C$= +60 V), and the concomitant FeRh unit cell expansion/contraction is sketched. Blue and white indicate FM and AFM regions, respectively. (d) Relative magnetic moment increase for V$_{pol}$ from -100 to 60 and back to -100 V measured following the arrow direction. e) XMCD-PEEM images at V$_{bias}$ = -100 V for a representative region of the sample. (f) XMCD (XMCD= (I-I$_{MAX}$)/I$_{MAX}$) versus voltage contrast of the signaled regions in (e). Reproduced with permission from ref. [56].

Fina et al. [54] reported on the time-dependent change of magnetization $\Delta M$ of FeRh films grown on piezoelectric PMN-PT substrates, after application of voltage V(t) pulses. Prior to measurement, the initial state of the FeRh/PMN-PT sample was prepared by prepoling with a large electric field (V$_{prepol}$ = -100 V) that sets the initial value of the magnetization of the sample. In the example of Figure 8(a) this would correspond to the low magnetization state. Next the V(t) pulse of smaller V > 0 amplitude was applied and the magnetization changed recorded under zero field. It turned out that, under zero magnetic field and in the metastability region around the phase transition, a V(t) pulse of suitable amplitude promotes an increase of magnetization $\Delta M$ of the film. Although at first sight this finding may seems obvious as the reduction of the piezo-strain under V > 0 (|V| << V$_{prepo}$ and |V| smaller than the coercive field of PMN-PT) favors FM phase as discussed above, the observation of an increase of magnetization at remanence implies that the newly formed FM regions are oriented in a given direction rather than randomly, as it could be naively expected in zero magnetic field. This suggest that preexisting FM nuclei dictate the magnetization direction of the newly strain-promote FM regions. Indeed, XMCD-PEEM images collected under V-bias appear to confirm this view [Figure 9(a,b)].

Moreover, it was found that for a certain range of V (V larger than the coercive field of the piezoelectric substrates), the induced magnetization is time dependent [$\Delta M(t)$]. Indeed, $\Delta M(t)$ displays a pronounced maximum for a short time, before dropping back to zero (recovering the initial magnetization state) for a longer time. *In-operando* X-ray diffraction data [Figure 9(c)] suggested that this behavior is due to the time evolution of the relative abundance of FM/AFM regions dictated by the piezo-stress [Figure 8(d)].

The observation that strain- promoted FM domains retain the magnetization direction of residual pre-existing FM regions, was also demonstrated by performing magnetization measurements at remanence after consecutive electric cycling in FeRh/PMN-PT and FeRh/BTO [55]. It was show that this approach allowed to write magnetic information in the FM state and thus cloak it by a suitable voltage pulse that brings the sample into the AFM. Eventually, data can be retrieved by a voltage pulse of inverse polarity that recovers the FM state. This idea was further extended and exploited by Fina et al. [56] to demonstrate the possibility of a reversible and magnetically unassisted voltage-driven switching of the magnetization in FeRh/PMN-PT heterostructures [Figure 10(a-d)]. It is remarkable that the large changes of magnetization can observed, under zero magnetic field, simply by applying a voltage stimulus [Figure 10(a)] and that magnetization change remains even when the strain pulse is suppressed [Figure 10(b)]. Applying suitable applied voltage excursion different non-volatile magnetic states can be obtained [Figures 10(d)]. In the obtained lower magnetization state the FeRh film is in-plane compressed and thus favoring the AFM state; and the higher magnetization state the FeRh films is in-plane less compressed and thus favoring the FM state. Local magnetization measurements by using XMCD-PEEM confirmed the reversible switching of the between AFM and FM domains by voltage induced strain change [Figure 10(e,f)]. These experimental results illustrate the pivotal role of pinned FM domains within the AFM matrix, guiding the magnetic polarization of the strain-controlled AFM- to FM transition, and dictating the direction of the remanent magnetization.

## 7. Summary and conclusions

Probably triggered by its potential applications in spintronics, the progress on fabrication and understanding the properties of FeRh films has been tremendous. Epitaxial films of high quality have been grown on various substrates allowing a basic understanding of the role of strain on the magnetic and electric properties. However, the hallmark has been the achievement of epitaxial grown on piezoelectric substrates that has allowed obtaining large modulation of magnetic and electric properties under modest electric fields. The arrested nucleation and growth of FM domains within the AFM phase and vice versa, under strain controlled AFM-FM transition, has allowed to achieve reversible and unassisted voltage-driven switching of the magnetization that may find opportunities in data storage. In spite of the outstanding progress, still many questions remain open. For instance, the role of interfaces, either with substrates and capping layers, which are expected to be increasingly larger when reducing film thickness and device size, although largely explored, still requires concerted actions of thin film preparation and exhaustive characterization to delineate the origin and impact of the prevalent FM regions observed at interfaces. On the other hand, it has been shown that strain allows modulating the saturation magnetization of the film, but there is a deficit of information in reference to magnetic anisotropy. The reported ability of FeRh films to store and hide information in the AFM state without assisting magnetic fields, links FeRh to current trends in AFM spintronics, where it may find new opportunities. Electric field control of magnetic and transport properties, successfully explored using ferroelectric single crystalline substrates of ferroelastic materials, has permitted to amplify the genuine piezoelectric strain effects at the cost of irreversibility. It could be of interest to explore non-ferroelastic substrates or thin film templates for the growth of FeRh aiming at large piezoelectric-controlled strain while preserving reversibility (not involving polarization switching). Last, development of FeRh on sacrificial substrates may allow developing a new generation of gauges and sensor of potential applications beyond spintronics. All in all, there is plenty of room to develop and exploit the intriguing properties of an alloy about 80 years old.

## Acknowledgements

Financial support from the Spanish Ministry of Economy, Competiveness and Universities, through the "Severo Ochoa" Programme for Centres of Excellence in R&D (SEV-2015-0496), MAT2015-73839-JIN (MINECO/FEDER, UE) and the MAT2017-85232-R (AEI/FEDER, EU) and from Generalitat de Catalunya (2017 SGR 1377) is acknowledged. IF acknowledges Ramón y Cajal contract RYC-2017-22531.